\documentclass{revtex4}
\usepackage{amssymb}
\usepackage{amsmath}
\usepackage[usenames]{color}
\usepackage{mathrsfs}
\usepackage{euscript}
\usepackage{graphicx}
\usepackage{graphics}
\usepackage{bm}
\usepackage{color}
\usepackage{pdfpages}

\begin{document}

\begingroup

{\Large\bfseries\mathversion{bold}
~~~~~
 \par}%
\vspace{8mm}



# Orientational decomposition of molecular high harmonic emission in three dimensions

Limor S. Spector[1,2*], Maxim Artamonov[3], Shungo Miyabe[2,4], Todd Martinez[2,4], Tamar Seideman[3], Markus Guehr[2], and Philip H. Bucksbaum[1,2]

[1]Applied Physics Department, Stanford University, Stanford, CA 94305, USA
[2]Stanford PULSE Institute, SLAC National Accelerator Laboratory, Menlo Park, CA 94025, USA
[3]Department of Chemistry, Northwestern University, Evanston, Illinois 60208, USA
[4]Chemistry Department, Stanford University, Stanford, CA 94305, USA

**An important goal in molecular physics and chemistry today is to obtain structure-dependent information about molecular function to obtain a deeper understanding into chemical reactions[1-4]. However, until now, asymmetric tops, which comprise the widest and most general class of molecules, remain principally unexplored. This gap is particularly evident in high harmonic generation (HHG). HHG has successfully obtained structural information about electron hole pairs[5] or orbitals[6] for simple linear molecules. Unfortunately, for more complicated molecules, the emission from different molecular directions interfere, concealing individual angular signatures. Here we introduce a method to extract orientation-dependent information from asymmetric tops and apply it to the sulfur dioxide ($SO_2$) molecule. We use the rotational revival structure to decompose the angular contributions of HHG emission. This method also extends HHG-based tomographic imaging into three dimensions and makes it applicable to a much wider class of systems than previously envisioned. Our results suggest that HHG is a powerful tool to probe electron orbital structure and dynamics of complex molecules.**

HHG has shown promise as a technique for understanding electron structure and dynamics. Since HHG is a phase-matched process, the signal it produces is a result of a large aggregation of molecules rotating at different speeds in different directions. This results in a flat signal over time unless the molecules are aligned[7]. Once aligned, molecules can be probed at different angles to obtain structural information.

This works well for molecules with only one relevant dimension since all molecules are induced to rotate along that dimension. But for asymmetric top molecules, which have three different moments of inertia, a linearly polarized alignment pulse leaves the molecules free to undergo uncontrolled rotation about two of the three axes. As shown in the upper portion of Fig. 1, the resulting signal obscures the original orientation information even while separating the harmonics by energy. This major limitation has restricted HHG-based imaging to only small atoms or linear molecules[5,8-10].

Here, we present an orientational decomposition method for obtaining both the angle-dependent amplitude and phase of harmonic emission in the molecular frame. Using the quantum revivals as spectral markers, we fit single-axis alignment patterns to our data to decompose the HHG signal. As depicted in the lower portion of Fig. 1, our analysis allows us to reconstruct the fundamental harmonic emissions by separating each harmonic's energy into three signals, one per axis.

Our method is generalizable to any molecule that can exhibit revivals. This result suggests a new approach to understanding the rotations of complex polyatomic

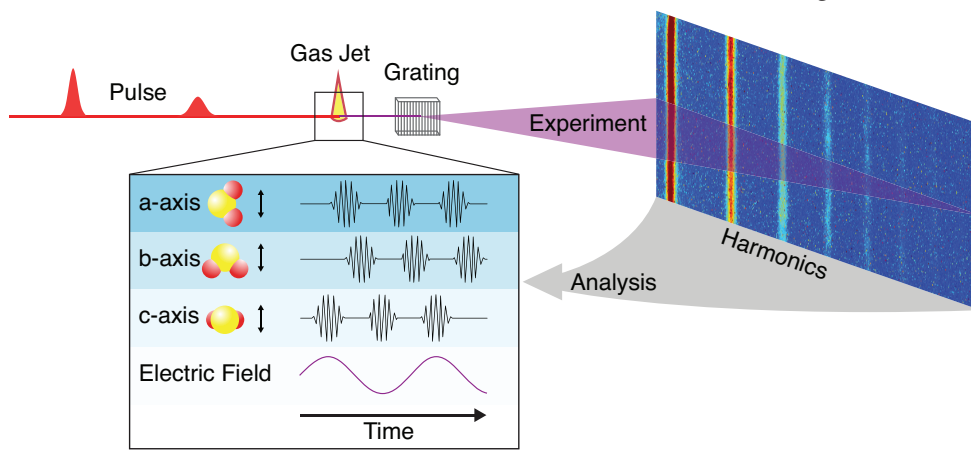

FIG. 1 The orientational decomposition method. For asymmetric top molecules, HHG emission from each molecular direction interferes and orientation-dependent information becomes obscured in the signal (top). Using this method, we fit single-axis alignment patterns to our data to decompose the HHG signal into the amplitude and phase from each molecular axis (bottom).

1

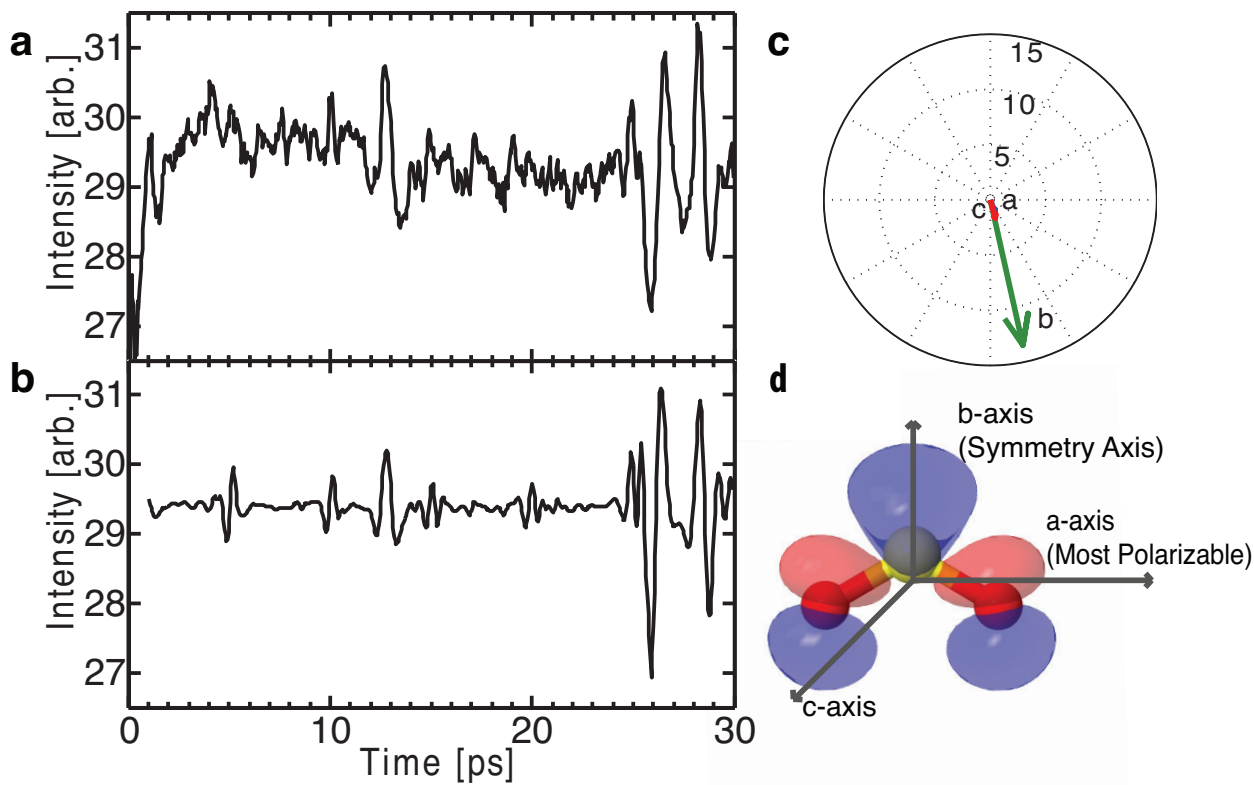

FIG. 2 Comparison between experiment and a best-fit theory curve. (a) Data for high harmonic 19 (29.5 eV) for the first 30 ps following alignment. The high harmonic data is very sensitive to SO2 revivals, with order up to 1/16 revivals visible. (b) Best-fit theory curve fits all 3 single-axis alignment patterns to the data, yielding orientational information. (c) A phasor plot representing the magnitude and direction of HHG emission as calculated by the best-fit curve. (d) A sketch of the HOMO of SO2, shown in red and blue to encode the phase, overlaid with a ball and stick model. We see highest HHG signal originating from the symmetry axis direction, in confirmation of our analysis.

molecules and hence their structural properties. Most molecules are asymmetric tops and even reactions with linear molecules generally proceed via an asymmetric top transition state. Thus, an orientational decomposition into three dimensions is a solution that would benefit much of the ultrafast science community, including users of techniques like time-resolved x-ray[11,12] and electron diffraction[13,14], photoelectron spectroscopy[15,16], and high harmonic generation (HHG)[17-19].

As a test case for our method we chose the asymmetric top $SO_2$. The experimental scheme employs two pulses: one for alignment and one probing HHG pulse. We vary the time delay between the two pulses to record the transient alignment revivals. Since the $SO_2$ molecules are aligned with respect to the field polarization vector at a molecular revival, the harmonic signal is sensitive to the molecular structure along the aligned axis. shown in Fig. 2a. The molecule shows revivals about all axes. We observe fractional revivals and multiple revivals to a sensitivity seen by high harmonic generation only in linear molecules so far[20]. Moreover, we observe all types of prolate-type asymmetric top rotational revivals, which have not previously been seen (see Supplementary Information)[21].

A few significant structures are immediately apparent in Fig. 2a: there is an initial alignment peak immediately following the impulse with a duration of less than 1 ps. A revival peak appears around 13 ps and a large and complex triply-peaked revival structure appears around 26 ps. These peaks, in addition to smaller ones, serve as distinguishing marks of the spectrum that we can use to decompose the signal into individual molecular axis contributions.

To aid in the decomposition, we make use of calculations of the single-axis alignment patterns for $SO_2$ (Fig. 3). The single-axis alignment patterns are the set of molecular alignments as viewed from a particular axis of the molecule (see Methods). The pattern is the expectation value $<\cos^2\theta_i>$ in the rotational wavepacket, where $\theta_i$ is the angle between the polarization direction of the HHG probing pulse and the molecular axis i. Each axis is sensitive to a multitude of revival types and so there is no simple correspondence between a particular axis and a particular revival type. The molecular alignment is calculated with respect to all three molecular axes, where a, b, and c are the molecular axes defined in Fig. 3h. An

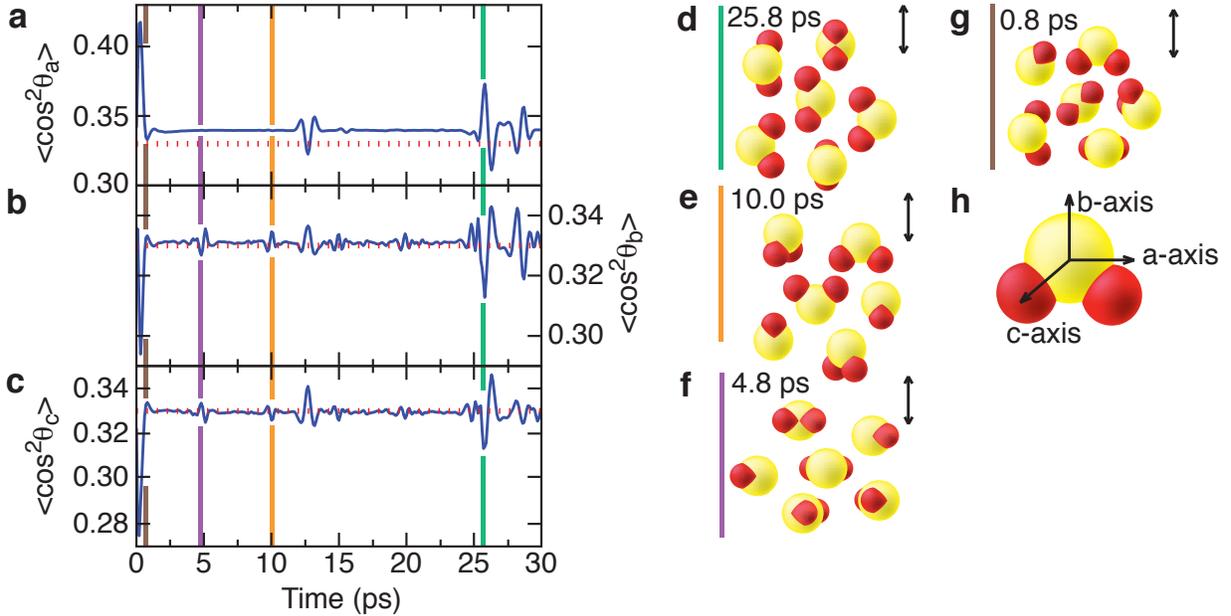

FIG.3 Single-axis alignment patterns. (a), (b), and (c) Expectation values of alignment between molecular directions a, b, and c and the polarization direction of the HHG probing pulse. The molecules are in an isotropic distribution for $\langle\cos 2\theta\rangle = 1/3$. (d), (e), (f), and (g) Idealized sample distributions at different revival times. The laser polarization direction is indicated. (d)-(f) correspond to alignments along axes a, b and c, respectively, which (g) corresponds to an isotropic distribution. (h) Molecular axes for $SO_2$.

isotropic ensemble has an expectation value of 1/3 as depicted by a dashed red line in the figure.

We can use the single-axis alignment patterns as basis vectors along which the data is decomposed. To extract the harmonic signal along each of the principal axes of the molecule, we combine the single-axis alignment patterns to fit the data to the form $\left|\sum_j c_j f_j\right|^2$, where $f_j$ are the three single-axis alignment patterns and the $c_j = |c_j| e^{i\theta_j}$ are complex coefficients whose amplitudes $c_j = |c_j|$ and phases $\theta_j$ serve as six Nelder-Mead fit parameters. The result is shown in Fig. 2b. Physically, the three coefficients can be viewed as the radiated electric field components along the three molecular axes. Importantly, this method allows us to extract separately the magnitude and phase of the harmonics.

The extracted fit coefficients for the magnitude of the electric field are $1.2^{+2}_{-2} : 13.5^{+2}_{-2} : 1.7^{+2}_{-2}$ for axes a:b:c on an arbitrary scale. For the phase the fit coefficients are $5.1^{+2}_{-.7} : 4.9^{+.5}_{-.3} : 5.0^{+.4}_{-.5}$ radians. The $\chi^2$ for the fit is 0.56 with 120 degrees of freedom (see Supplementary Information).

The fit coefficients are shown as a phasor diagram in Fig. 2c. We conclude that emission comes primarily from molecules aligned with the polarization along the $C_2v$ (b) axis. We can see that, for this harmonic, the dominant contribution to high harmonic emission is in the b-axis direction. There is some emission along the a- and b- directions as well, but their magnitude is small compared to the b-axis. Moreover, the phase is approximately in the same direction for all three axes, so the b-axis remains the dominant contribution. The phase is sensitive to emission and recombination[17]. Our data further indicate that the harmonic emission is suppressed at the orientations along the a-axis, where the polarizability is largest.

The high contribution to harmonic emission from the b-axis could be due to either enhanced field ionization or enhanced recombination along this axis, according to the standard recollision model[22-24]. We model the energy- and angle-dependent recombination dipole by calculating the VUV photoemission cross section of $SO_2$ in the molecular frame, as shown in Fig. 4.

We find that the recombination dynamics are the ones to predominantly determine the direction of harmonic emission in the molecular frame. At our signal energy of 29.5 eV there is a large recombination probability along the molecular symmetry axis, but much less recombination along the other two axes. This is consistent with the results of our fit, and provides a second confirmation of its validity. We compared our results from the HOMO to recombination probabilities arising from the other valence orbitals (HOMO-1 through HOMO-5), as shown in Fig. S1. The recombination dipole dependence on geometry for these orbitals shows that our signal arises primarily from the HOMO.

We can understand the recombination dynamics by

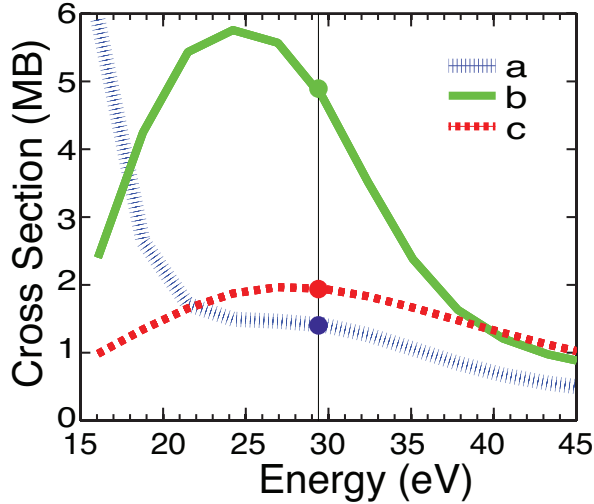

FIG. 4 Recombination cross section of SO2 as a function of photon energy. The laser polarization is oriented along the a, b and c axes of the molecule in the dotted blue, solid green, and dashed red curves, respectively. High harmonic 19 has approximately 29.5 eV of energy; at this point the recombination cross section is much higher along the molecular symmetry axis (b-axis) than along either of the other two axes.

als confirms the sensitivity seen by Lock et al. for HHG[20], and opens HHG as a viable medium for studying alignment even in asymmetric top molecules. Second, we show that the results of this method may reveal orbital structure. This creates many interesting avenues for further study. For example, by comparing the results of this method seen by different harmonic spectra, we may be able to learn about strong-field tunneling from the electron orbitals.

**Methods Summary:**

*Experiment*

We use impulsive stimulated rotational Raman scattering for alignment with one 130 fs pulse at 800 nm with an intensity in the 1013 W/cm² range to aligning a sample of $SO_2$ molecules. Following the alignment pulse, the $SO_2$ is free to rotate, exhibiting a revival structure that is determined by the field-free rotational spectrum. We probe the $SO_2$ at various time delays by using a focused 30 fs pulse at 800 nm with an intensity of approximately $2.6 \times 10^{14}$ W/cm² to produce HHG in the $SO_2$ gas. Both laser beams are linearly polarized parallel to each other. We vary the time delay between the two pulses between 0 and 30 ps. The HHG is dispersed through a flat-field diffraction grating spectrometer and amplified with a microchannel plate and detected on a phosphor screen. The molecules are introduced through a nozzle located above the laser focus and thus slightly cooled to $240 \pm 20°$ K at the laser focus.

*Theory of single-axis alignment patterns*

The aligning pulse duration and intensity in the calculation are taken to be the same as used in our experiments. The observables are calculated for a thermal ensemble corresponding to the temperature of 180° K, the highest temperature that was computationally feasible. Our theory and numerical approach for solving the time-dependent Schrödinger equation of an asymmetric top molecule interacting with an aligning field are described in detail in[25] and[7]. In brief, for the duration of the laser pulse, the rotational wavefunction is expanded in a symmetric top basis, |JKM>, thus converting the time-dependent Schrödinger equation into a set of coupled differential equations that are solved numerically. Here J, K, and M denote the quantum numbers corresponding to the total angular momentum and its projections onto the body- and space-fixed z-axes. Because the aligning pulse is linearly polarized, the wavefunction is effectively two-dimen-

examining the orbital structure for $SO_2$ shown in Fig. 2d. Since there is greater electron density along the symmetry axis there is a higher recombination probability in that direction than along the other two directions, even than along the most polarizable axis direction. The analysis suggests that the orientation of harmonic emission is largely controlled by the recombination dynamics and hence responds directly to the form of the electronic orbital that dominates recombination.

The method we have presented here thus provides the first way to decompose the orientational information from HHG into the molecular frame basis. This orientational information is crucial for determining structure and transient structure in asymmetric tops. One advantage of this method is that it uses only a single impulse to extract information about all three molecular axes. It removes the necessity to align a polyatomic molecule along all three dimensions, an onerous requirement that places a burden on the amount of laser power needed. Another advantage is that all of the high harmonic information – magnitude and phase – is determined completely experimentally, with no model-dependent information. We expect that orientational decomposition will be useful for other studies requiring alignment as well.

Here we concentrate on harmonic generation along the principal axes of the model only, and so we stop short of complete tomographic reconstruction. However, this same decomposition along principal axes could be repeated along any polarization angle with respect to the driving pulse polarization for a full reconstruction. It would be very interesting to do a full tomographic reconstruction on an asymmetric top, because the high harmonic generation should have complicated relationship to molecular structure, tunneling and interferences. Our method thus provides a path towards HHG-based tomographic reconstruction of asymmetric top molecules.

There are two other implications of our results. First, our identification of revival features up to the 1/16 reviv-

sional with only parametric dependence on M. Note that K is not a conserved quantum number for an asymmetric top. The pulse is modeled as a Gaussian. After the pulse turn-off, when the envelope tail is truncated as it becomes sufficiently small, the wavefunction is transformed to the basis of eigenstates of the field-free Hamiltonian. The principal components of the polarizability tensor were determined in[26].

*Photoionization calculations*

We computed fixed-nuclei photoionization amplitudes using the complex Kohn variational method[27]. This gives us information about the recombination physics, because the recombination step in HHG is equivalent to inverse photoionization. The final-state electronic wavefunction for production of photoions is in a specific cation state. Only one ionic target state, the $8a_1^{-1}$ hole state, was included in the trial wave function. Further details can be found elsewhere[28,29] (see Supplementary Information). Photoionization cross sections in the molecular frame were then constructed from the matrix elements

$$I^\mu_{\Gamma_0 l_0 m_0} = <\Psi_0 | r^\mu | \Psi_{\Gamma_0 l_0 m_0}>,$$

where $r^\mu$ is a component of the dipole operator, which we evaluate here in the length form. $\Psi_0$ is the initial-state wave function of the neutral N electron molecule.

 

{\Large\bfseries\mathversion{bold}

Orientational decomposition of molecular high harmonic emission\\ \indent in three dimensions \par}%
\vspace{8mm}

\begingroup\scshape\large
\emph{Supplementary Information}
\endgroup
\vspace{12mm}

\begingroup\scshape\large\indent
Limor S. Spector$^{1,2*}$, Maxim Artamonov$^{3}$, Shungo Miyabe$^{2,4}$ Todd Martinez$^{2,4}$,\\
\indent Tamar Seideman$^{3}$, Markus Guehr,$^{2}$ and Philip H. Bucksbaum$^{1,2}$\\
\endgroup
\begingroup\small

$^{1}$ \emph{Applied Physics Department, Stanford University, Stanford, CA 94305, USA} \\
\indent $^{2}$ \emph{Stanford PULSE Institute, SLAC National Accelerator Laboratory, Menlo Park, CA 94025, USA}\\
\indent $^{3}$ \emph{Department of Chemistry, Northwestern University, Evanston, Illinois 60208, USA} \\
\indent $^{4}$ \emph{Chemistry Department, Stanford University, Stanford, CA 94305, USA}\\
\endgroup

\section{Analyzing the harmonic signal}

To generate HHG, we focused the output of a commercial Ti:Sapphire laser with a pulse duration of  about 30 fs, a pulse energy of about 53 $\mu$J and a central wavelength of about 780 nm onto a continuous flow gas jet using a focusing lens with f = 150 mm.  The alignment pulse was made from the same laser, but the pulses were chirped to a 130 fs pulse width using 16 mm of BK7 glass, and retained a pulse energy of about 28 $\mu$J.  The HHG pulse followed the alignment pulse through the gas jet, and harmonics that were produced in jet and were between 25 and 55 eV passed through an aluminum filter with a thickness of 100 nm onto a spherical grating.  The experimental scan was taken in steps of 47.8 fs.  The dispersed image was captured by an extreme ultraviolet detector and image intensifier consisting of a bare microchannel plate (MCP) followed by a phosphor screen.  This was then imaged using a charge-coupled device (CCD) camera.

Before analyzing the high harmonic signal, we accounted for several effects, as is common in HHG, see, for example Shiner \emph{et al.}~\cite{Shiner}.  The spherical grating focuses only in the dispersion direction in the incidence plane of the EUV light (tangential direction) but keeps the natural divergence of the harmonic beam in the orthogonal (sagittal) direction. Since the beam hits the grating under grazing incidence, the finite size of the grating substrate acts like a slit that filters only the center part of the beam in the tangential direction. We accounted for the wavelength transmission function of the apparatus in the data we present.  We measured the wavelength-dependent transmission of the aluminum filter in our laboratory in a separate experiment, which takes the real oxidation of the filter into account. The wavelength transmission of the grating is provided by the manufacturer (Hitachi) and we estimated the MCP efficiency from Hemphill and Rogers~\cite{Hemp}.

We chose harmonic 19 for this study, because we wanted to concentrate on a harmonic that was intense and thus more immune to statistical noise, but at the same time also in the high harmonic plateau region and far from the cutoff, where the results are strongly dependent on intensity.  We also wanted a harmonic that would be far from the lower order harmonics where the standard three-step model is less valid.  The fit for neighboring harmonics reveal can reveal additional information about the energy dependence of the harmonic process, and we intend to investigate this more fully in future work.

We estimated the temperature by assuming a supersonic expansion.  We calculated the backing density, $2.43\times10^{19}/cm^{3}$, by using the equation of state for an ideal gas with a backing pressure of 1 bar and room temperature.   Our uncertainty stems mainly from our lack of precise knowledge of how far exactly the high harmonic jet is from the laser focus, x, and precisely how large our tip diameter, d, is.  Our best estimate for this parameter is $.033<x/d<1$.  In Scoles \emph{et al.}~\cite{Scoles}, the authors show that for a supersonic gas jet,
\begin{equation}
  \label{eq:Scoles1}
\left({\frac{T}{T_{0}}}\right)^\frac{1}{\gamma-1}=\left(1+\frac{\gamma-1}{2}M^2\right)^{-\frac{1}{\gamma-1}}
 \end{equation}
where $T_0$ represents the initial temperature,  $T$ represents the final temperature,  $\gamma$ is the heat capacity ratio (1.29 for SO$_{2}$) and $M$ is the Mach number for the supersonic expansion.  $M$ is calculated by
 \emph{et al.}~\cite{Scoles}, the authors show that for a supersonic gas jet,
\begin{equation}
  \label{eq:Scoles2}
 M=1.0+A{\left(\frac{x}{d}\right)}^2+B{\left(\frac{x}{d}\right)}^3
 \end{equation}
 for $0<x/d<1$.  A and B are the Mach number correlation parameters for axisymmetric expansion and are given by A = 3.61 and B = 1.95.  From these equations we were able to estimate the temperature of the gas jet at the laser focus as 240 $\pm$ 20$^\circ$K.

 \section{Revivals in Asymmetric Tops}
 
Revival is a standard term of art developed for symmetric tops such as diatomic molecules, which we adopt here for the more complex case of asymmetric tops following rotational Raman excitation by a short polarized laser pulse. The basic model remains the same in that a polarized light pulse impinges on a gas phase sample, which experiences initial alignment and then dephasing and rephrasing. The sample experiences a rephrasing, because the molecules rotate at frequencies that are multiples of the three rotational constants. In other words, the revival refers to a periodic peak in the alignment expectation value, and is a pure quantum feature related to the fact that quantized rotational levels form a non-equidistant progression.  This is the reason why 3-dimensional revivals of an asymmetric top have been so elusive.  Our method resolves this problem by not trying to relate a principal axis to any particular revival time, but rather relating each axis to a distinct pattern of alignment revivals utilizing the full rotational spectrum.

In a diatomic molecule the rotational revivals occur at the times when the molecular axes are optimally aligned, but the situation in an asymmetric top is more complicated.  While linear and symmetric top molecules exhibit complete periodic revivals, asymmetric top molecules undergo classically unstable motion and hence do not exhibit complete reconstruction of the initial state.  The three axes of an asymmetric top molecule, namely a, b, and c, arise from the rotational constants A, B and C of the molecule itself. (A, B and C are inversely proportional to the moments of inertia of the molecule.) Since the different types of rotational coherence effects are determined by the superposition of states that caused them, there is no one-to-one correspondence between the axis labels, a, b, and c, and the revival types. 

To understand and characterize the alignment of the molecular ensemble as a function of time it is instructive to simultaneously consider the three curves in Fig. 3 in the main text.  A peak in a given curve indicates preferential alignment along the corresponding axis.  For the maximum at 25.8 ps, for instance, the molecules are preferentially aligned along the axis with the peak.  For 25.8 ps, a sketch of an ensemble is shown in Fig. 3d, with the molecules aligned along the a-axis.  Other distributions at different times are shown in Fig. 3e-g, showing that the molecules preferentially align along each axis at different times.

The interference of selection rules with J-levels and K-levels characterize each revival type, where J and K refer to the symmetric top quantum numbers. Although K is no longer a good quantum number for an asymmetric top molecule due to the mixing of K-states, in the limit of an oblate-type or prolate-type asymmetric top molecule, we can refer to K as the symmetric top equivalent. The rotational coherence types are shown in Table \ref{tab:rotationalTypes}.
 
 \begin{table}[!t]
\centering
\begin{tabular}[t]{c|c|c}
\hline\hline
 \textrm{Transition Type} &\textrm{Revival Type} & \textrm{Contributing Coherences}\\
\hline
J-type	& P: $t\approx\frac{n}{2(B+C)}$ & $|\Delta J|=1,2$, $|\Delta K|=0$\\
~ & O: $t\approx\frac{n}{2(A+B)}$ & \\
K-type	& P: $t\approx\frac{n}{4A-2B-2C}$ & $|\Delta J|=0$, $|\Delta K|=2$\\
~ & O: $t\approx\frac{n}{|4C-2A-2B|}$ & \\
Hybrid	& P: $t\approx\frac{n}{2A-B-C}$ & $|\Delta J|=0$, $|\Delta K|=1$\\
~ & O: $t\approx\frac{n}{|2C-A-B|}$& \\
C-type & $t\approx\frac{n}{4C}$ & P: $|\Delta J|=2$, $|\Delta K|=0$\\
~ & & O: $|\Delta J|=2$, $|\Delta K|=2$\\
A-type &$t\approx\frac{n}{4A}$	& P: $|\Delta J|=2$, $|\Delta K|=2$\\
~ & & O: $|\Delta J|=2$, $|\Delta K|=0$\\
\hline\hline
\end{tabular}
\caption{ Rotational coherence types for asymmetric top molecules.  P and O refer to prolate-type and oblate-type molecules, respectively, and $n$ refers to the revival order.}
\label{tab:rotationalTypes}
\end{table}

The different types of rotations in SO$_2$ that are responsible for revivals are conveniently characterized using the nomenclature developed in rotational coherence spectroscopy~\cite{Poulsen}, although our present strong field, far-off-resonance excitation gives rise to qualitatively different rotational wavepackets.   Asymmetric tops are classified as near-oblate or near-prolate in analogy with symmetric top molecules.  As a near-prolate asymmetric top molecule, SO$_2$ is expected to exhibit J-type and K-type revivals corresponding to quasi-symmetric-top-like rotations.  Since SO$_2$ is, however, an asymmetric top, it can also exhibit A-type, C-type and Hybrid-type revivals. 

SO$_2$ is a prolate-type asymmetric top molecule and thus follows the ÒPÓ revival times referred to in Table \ref{tab:rotationalTypes}. Table \ref{tab:expectedRev} tabulates some of these revivals for SO$_2$.  The extensive results of rotational coherence spectroscopy allow us to compute the revival times for SO$_2$~\cite{Felker}.  Looking closely at Fig. 2a in the main text, we can identify partial and full J-type revivals at around 13 and 26 ps, respectively, a C-type revival at 28 ps, and several A-type, K-type, and Hybrid-type revivals at multiples of 4, 5 and 10 ps, respectively, except for where they are masked by the partial revival signature.  We also see other partial revivals, of order up to 1/16 revivals.  Some of the revivals that can be seen in the data are marked in bold in Table \ref{tab:expectedRev}.
 
\begin{table}[!t]
\centering
\begin{tabular}[t]{c|c|c|c|c|c}
\hline\hline
\textrm{~} &  \textrm{A-type} &\textrm{C-type} & \textrm{J-type} & \textrm{K-type} & \textrm{Hybrid-type}\\
\hline
1/4	& 1.0 & \bf{7.1} & \bf{6.5} & 1.2 & 2.4\\
1/2	& \bf{2.1} & \bf{14.2} & \bf{13.1} & \bf{2.4} & 4.9\\
3/4	& 3.1 & \bf{21.3} & \bf{19.6} & \bf{3.7} & 7.3\\
 & & & & & \\
Full	 & \bf{4.1} & \bf{28.4} & \bf{26.2} & \bf{4.9} & \bf{9.8}\\
1 1/4 & 5.1	& 35.5 & 32.7 & \bf{6.1}	& 12.2\\
1 1/2	& \bf{6.2} & 42.6 & 39.2 & \bf{7.3}	& 14.6\\
1 3/4	& 7.2	& 49.7	& 45.8	& \bf{8.5}	& 17.1\\
  & & & & & \\
2nd	& \bf{8.2}	& 56.8	& 52.3	& \bf{9.8}	& \bf{19.5}\\
2 1/4	& 9.3 & 63.9	& 58.8	& \bf{11.0}	& 22.0\\
2 1/2	& \bf{10.3}	& 71.0	& 65.4	& 12.2	& 24.4\\
2 3/4	& 11.3	& 78.1	& 71.9	& 13.4	& 26.8\\
   & & & & & \\
3rd	& \bf{12.3}	& 85.2 & 78.5 &\bf{14.6}	& \bf{29.3}\\
3 1/4	& 13.4	& 92.3	& 85.0	& \bf{15.9}	& 31.7\\
3 1/2	& 14.4	& 99.4	& 91.5	& \bf{17.1}	& 34.2\\
3 3/4	& 15.4 & 106.5	& 98.1	& 18.3	& 36.6\\
    & & & & & \\
4th	& \bf{16.5}	& 113.6	& 104.6	& \bf{19.5}	& 39.0\\
4 1/4	& 17.5	& 120.7	& 111.1	& \bf{20.7}	& 41.5\\
4 1/2	& 18.5	& 127.8	& 117.7	& \bf{22.0}	& 43.9\\
4 3/4	& 19.5 & 134.9	& 124.2	& \bf{23.2}	& 46.4\\
     & & & & & \\
5th	& \bf{20.6} & 142.0	& 130.8	& 24.4	& 48.8\\
      & & & & & \\
6th	& \bf{24.7} & 170.5	& 156.9	& \bf{29.3}	& 58.6\\
\hline\hline
\end{tabular}
\caption{Some expected revivals for SO$_2$ (in ps)}.  Bold face revival times indicate that they can be seen in the data.
\label{tab:expectedRev}
\end{table}
 
\section{Developing the fit}

To preserve the simplicity of the model we chose a straightforward method for fitting.  To determine the coefficients used for the fit shown in Fig. 2b, we minimized the $\chi^2$ of a six-parameter Nelder-Mead fit.  The six parameters of the fit correspond to the magnitude and phase components of the dipole moment along the three axis directions.  The $\chi^2$-based formula used for minimization is:

\begin{equation}
  \label{eq:chi2}
 \frac{\left(d-|{\sum_{j}c_jf_j|^2}\right)^2}{|d|}
 \end{equation}where the $c_j$ are the three fit coefficients, the $f_j$ correspond to the three single-axis alignment patterns and $d$ corresponds to high harmonic data.   The three coefficients can thus be viewed as corresponding to radiated electric field components.  This method allows us to extract separately the electric field magnitude and phase by allowing each $c_j$ to be of the form $|E|e^{i\theta}$.  Both d and f are summed over all data points.  

The combined theory curve is able to capture the most prominent features of the experiment, like the J-type and C-type revivals.  On a more detailed level, since the theory does not capture all of the revival times, combining all three single axis revival curves into one complete theory curve will still not capture all of the revivals.  Therefore we do not expect Fig. 2b to ever look entirely like Fig. 2a with this analysis.  However, we do not need the theory to capture all of the revivals.  Since the main features, namely the J-type and C-type revivals, are captured by the theory, we can successfully use a fit to match the data and find the amount that each axis will contribute to the overall harmonic emission.

$\chi^2$ for the fit is 0.56 for 120 degrees of freedom.  The critical $\chi^2$ statistic for a significance level of $\alpha$ = .05 and 120 degrees of freedom is 147.  Since $0.56<147$, we must accept the null hypothesis that the fit matches the data.  To ensure that this result was not dependent on the number of degrees of freedom used, we calculated $\chi^2$ using different binnings. The results are tabulated in Table \ref{tab:chi2Table} below, and are comparable.

We expect harmonic generation to have a nonlinear dependence on geometry since the field-ionization step in the standard semi-classical model of the process is very sensitive to small changes in the charge density at the laser-induced saddle point in the potential.  The process is also sensitive to the recombination dipole moment.  A more complete model would include in the density matrix vibrational and rotational excitations produced by the impulse.  Although this may affect close correspondence of fit to small details in the data, this does not affect the information we extract from the fit, which are the locations and relative amplitudes of revivals. 

Our main conclusion from the fit is that most of the fitting weight is in on the single-axis alignment pattern $ÒbÓ$.  To this end, we also modeled the data using only the $ÒbÓ$ single-axis alignment pattern.  This yields a $\chi^2$ value of 54.4, which is much higher than our initial fit.  Although the b-axis alone is sufficient to model harmonic 19, using all three axes is yields a much better than using the b-axis alone.

To test this method and make sure that the b-axis alone does not describe every harmonic, we also examined harmonic 21.  For harmonic 21, we were able to make a fit with a $\chi^2$ value of 0.19 for with 120 degrees of freedom.  For a fit with the b-axis alone, we obtain a value of 486.  This is above the critical $\chi^2$ value of 147 for 120 degrees of freedom, and we must reject the null hypothesis, meaning that for the b-axis alone can not describe harmonic 21~\cite{Spector}.

The errors are produced by using the standard technique of keeping one parameter fixed while allowing the others to minimize $\chi^2$~\cite{NumericalRecipes}.  The errors thus measure the steepness of the minimum of $\chi^2$ along different directions.  Since the $\chi^2$ distribution is not symmetric, we obtain different upper and lower bound numbers.  

We minimized to $\Delta\chi^2$ = 6.63, which corresponds to 99\% degree of confidence assuming standard Gaussian statistics.  It is likely that in our experiment, as in others ~\cite{Pumplin, Nadolsky, Martin}, there are unknown nonnormal systematic errors.  These may cause the degree of confidence to be lower than stated above~\cite{Amsler}.  It is for this reason that we have quoted the 99\% degree of confidence rather than $\Delta\chi^2$ = 2.71 (90\%) or $\Delta\chi^2$ = 1 (68\%).

\begin{table}[!t]
\centering
\begin{tabular}[t]{c|c|c|c}
\hline\hline
\textrm{Degrees of Freedom} &\textrm{ $\chi^2_{ALL}$} & \textrm{$\chi^{2B}$} & \textrm{$\chi^2_{CRIT}$}\\
\hline
604 & 3.2 & 272 & 662\\
120 & 0.6 & 54 & 147 \\
54 & 0.2 & 25 & 72 \\
4 & 0.008 & 2.2 & 9.5\\
\hline\hline
\end{tabular}
\caption{ $\chi^2$ values using all axes and only the b-axis using different numbers of bins. The critical $\chi^2$ statistic is shown for a significance of $\alpha$ = .05.}
\label{tab:chi2Table}
\end{table}

\section{Theory for Single-Axis Alignment Patterns}

 Our theory and numerical approach for solving the time-dependent Schr{\"o}dinger equation of an asymmetric top molecule interacting with an aligning field are described in detail in Refs. \cite{SeidemanAAMOP06} and \cite{ArtamonovJCP08}. Here we give a brief outline of the procedure. The total Hamiltonian is $H(t)=H_\mathrm{rot}+H_\mathrm{ind}(t)$. The rotational Hamiltonian in the rigid-rotor approximation is
 \begin{equation}
   \label{eq:Hrot}
   H_\mathrm{rot}=B_X\hat{J}^2_X+B_Y\hat{J}^2_Y+B_Z\hat{J}^2_Z,
 \end{equation}
where $B_k$ and $J_k$, $k=X,Y,Z$, are the rotational constants components of the angular momentum operator, respectively. The body-fixed frame (BFF) is defined by setting the $X$-, $Y$-, and $Z$-axes parallel to the molecular $c$-, $b$-, and $a$-axes. In the space-fixed frame, the laser field polarization vector defines the $z$-axis. The field-matter interaction Hamiltonian is
\begin{equation}
  \label{eq:Hind}
  H_\mathrm{ind}=-\frac{\varepsilon^2(t)}{4}\left[\frac{\alpha^{ZX}+\alpha^{ZY}}{3}D^2_{00}-\frac{\alpha^{YX}}{\sqrt{6}}\left(D^2_{02}+D^2_{0-2}\right)\right],
\end{equation}
where $\varepsilon(t)$ is the aligning pulse Gaussian envelope, $D^2_{qs}$ are Wigner rotation matrices~\cite{Zare-book}, and $\alpha^{kk'}=\alpha_{kk}-\alpha_{k'k'}$ are the generalized polarizability anisotropies with $\alpha_{kk}$ being the BFF components of the polarizability tensor taken from Ref. \cite{LukinsJPC85}. The aligning pulse duration and intensity in the calculation are taken to be the same as used in our experiments. 

The rotational wavefunction is expanded in a symmetric top basis, $\{|JKM\rangle\}$,
\begin{equation}
  \label{eq:JKM}
  \langle\hat{R}|JKM\rangle=\sqrt{\frac{2J+1}{8\pi^2}}D^{J*}_{MK}(\hat{R}).
\end{equation}
Here $J$, $K$, and $M$ denote the quantum numbers corresponding to the total angular momentum and its projections onto the body- and space-fixed $z$-axes. The matrix elements of $H_\mathrm{rot}$ in this basis are easily evaluated analytically\cite{Zare-book}. Likewise, the matrix elements of $H_\mathrm{ind}$ are analytical and given as superpositions of integrals of the form
\begin{eqnarray}
  \label{eq:3j}
  \langle JKM|D^2_{0s}|J'K'M'\rangle&=&\delta_{MM'}(-1)^{K'+M'}\left[(2J+1)(2J'+1)\right]^{1/2}\times\\ \notag
&&\times\begin{pmatrix}J&2&J'\\
M&0&-M\end{pmatrix}\begin{pmatrix}J&2&J'\\
K&s&-K'\end{pmatrix}.
\end{eqnarray}
$H_\mathrm{rot}$ conserves $M$ quantum number. Note that $K$ is not a conserved quantum number for an asymmetric top. Because of the cylindrical symmetry of the linearly-polarized field, $M$ is also conserved by $H_\mathrm{ind}$ (see Eq. (\ref{eq:3j})). Because $M$ remains a good quantum number, and the wavefunction is effectively two-dimensional with only parametric dependence on $M$, i.e.,
\begin{equation}
  \label{eq:wf}
  |\psi^M(t)\rangle=\sum_{JK}C^M_{JK}(t)|JK;M\rangle.
\end{equation}
The time-dependent Schr{\"o}dinger equation is, thus, converted into a set of coupled differential equations to be solved numerically, i.e.,
\begin{equation}
  \label{eq:CDE}
  i\dot{C}^M_{JK}(t)=\sum_{J'K'}\langle JK;M|H(t)|J'K';M\rangle C^M_{J'K'}(t).
\end{equation}
After the pulse turn-off, when the envelope tail is truncated as it becomes sufficiently small, the wavefunction is transformed to the basis of eigenstates of the field-free Hamiltonian, $H_\mathrm{rot}$.

The alignment observables are $\langle\cos^2\theta_i\rangle$, $i=a,b,c$, where $\theta_i$ is the angle angle between the laser field polarization direction and molecular axis $i$,
\begin{equation}
  \label{eq:ta}
  \langle\cos^2\theta_{a}\rangle=\langle(\hat{z}\cdot\hat{Z})^2\rangle=\langle\frac{2}{3}D^2_{00}+\frac{1}{3}\rangle,
\end{equation}
\begin{equation}
  \label{eq:tb}
  \langle\cos^2\theta_{b}\rangle=\langle(\hat{z}\cdot\hat{Y})^2\rangle=\langle-\frac{1}{3}D^2_{00}-\frac{1}{\sqrt{6}}\left(D^2_{02}+D^2_{0-2}\right)+\frac{1}{3}\rangle,
\end{equation}
\begin{equation}
  \label{eq:tc}
  \langle\cos^2\theta_{c}\rangle=\langle(\hat{z}\cdot\hat{X})^2\rangle=1-\langle\cos^2\theta_{a}\rangle-\langle\cos^2\theta_{b}\rangle.
\end{equation}
The observables are calculated for a thermal ensemble corresponding to the temperature of 180 K,
\begin{equation}
  \label{eq:thermo}
  \langle O\rangle_T(t)=\sum_i\omega_i(T)\langle O\rangle_i(t),
\end{equation}
where the sum runs over all thermally populated initial states, $\langle O\rangle_i(t)$ is a state-specific observable (Eqs. \ref{eq:ta}--\ref{eq:tc}), and $\omega_i(T)$ are normalized weight functions consisting of the Boltzmann factor and nuclear spin statistical weight~\cite{RiehnPCCP05}.
 
\section{Computation of recombination cross sections}

Fixed-nuclei photorecombination amplitudes were computed using the complex Kohn variational method~\cite{rlm95}. Here we give a brief summary.  The initial-state wavefunction for the molecule in a specific {cation} state $\Gamma_0$ {and with initial angular momentum $l_0 m_0$} is written as
\begin{equation}
{\Psi_{\Gamma_0 l_0 m_0}=\sum_{\Gamma l m} A(\chi_\Gamma F_{\Gamma l m \Gamma_0 l_0 m_0}) +\sum_i d_i^{\Gamma_0 l_0 m_0} \Theta_i}
\end{equation}
where $\Gamma$ labels the  ionic target states $\chi_\Gamma$ included,  {$F$} are channel functions that describe the incoming electron, $A$ is the antisymmetrization operator and the $\Theta_i$'s are $N$ electron correlation terms. 
In the present application, only one ionic target state is included in the trial wave function, that being the $8a_1^{-1}$ hole state.

In the Kohn method, the channel functions are further expanded, in the molecular frame, as
\begin{equation}
\begin{split}
{F_{\Gamma l m \Gamma_0 l_0 m_0}}(\mathbf{r})  =  
& \sum_i {c_i^{\Gamma l m \Gamma_0 l_0 m_0}}\varphi_i(\mathbf{r}) \\
&+
\sum_{lm} 
\Big[
f_{lm}(k_\Gamma, {r})\delta_{ll_0}\delta_{mm_0} \delta_{\Gamma\Gamma_0} \\
&+
T_{ll_0mm_0}^{\Gamma\Gamma_0} h_{l m}^+(k_\Gamma, {r})
\Big]
Y_{lm}(\hat{\mathbf{r}})/{k_\Gamma^{\frac{1}{2}}r}\, ,
\end{split}
\label{eq:channelfcn}
\end{equation}
where the $\varphi_i(\mathbf{r})$ are a set of square-integrable (Cartesian Gaussian) functions,  {\it Y}$_{lm}$ is a normalized spherical harmonic, k$_{\Gamma}$ are channel momenta, and the $f_{lm}(k_\Gamma, {\mathbf r})$ and $h_{l m}^{+}(k_\Gamma, {\mathbf r})$ are numerical continuum functions that behave asymptotically as regular and outgoing partial-wave Coulomb functions, respectively~\cite{Resc}. The coefficients T$_{ll_0mm_0}^{\Gamma\Gamma_0}$ are the T-matrix elements.

Photorecombination cross sections in the molecular frame can be constructed from the matrix elements
\begin{equation}
{I^\mu_{\Gamma_0 l_0 m_0}=<\Psi_0|r^\mu|\Psi_{\Gamma_0 l_0 m_0}}>\, ,
\end{equation}
where $r^\mu$ is a component of the dipole operator, which we evaluate here in the length form, 
\begin{equation}
r^\mu = \Bigg\{
\begin{array}{ll} z, &
\mu=0\\
\mp \left(x\pm i y\right)/\sqrt{2}, & \mu=\pm 1 \end{array}\,
\end{equation}
and $\Psi_0$ is the final state wave function of the neutral $N$ electron molecule. In order to construct an amplitude that represents an electron with momentum ${\bf k}_{\Gamma_0}$ recombining with the molecule and ejecting a photon with  polarization direction $\hat{\epsilon}$, measured relative to the molecular body-frame, the matrix elements {$I^\mu_{\Gamma_0 l_0 m_0}$} must be combined in a partial wave series
\begin{equation}
\label{amplitude}
{I_{\hat{k} , \Gamma_0, \hat{\epsilon}}}=\sqrt{\frac{4\pi}{3}}\sum_{\mu l_0m_0} i^{-{l_0}}e^{i\delta_{l_0}} {I^\mu_{\Gamma_0 l_0 m_0}}
Y_{1\mu}(\hat{\epsilon})Y_{l_0m_0}{(\hat{k})}\, ,  
\end{equation}
where $\delta_{l_0}$ is a Coulomb phase shift. The cross section, differential in the angle of incoming electron and photon polarization relative to the fixed body-frame of the molecule, is then given by
\begin{equation}
\frac{d^2\sigma}{d\Omega_{\hat{k},{\Gamma_0}} d\Omega_{\hat{\epsilon}}}=
\frac{8\pi \omega}{3c}{|I_{\hat{k},{\Gamma_0},\hat{\epsilon}}|^2}\, ,
\end{equation}
where $\omega$ is the photon energy and $c$ is the speed of light. Finally the cross section for a particular polarization direction $\mu$ is given by
\begin{equation}
\sigma^{\mu,\Gamma_0}=
\frac{8\pi \omega}{3c}\sum_{l_0m_0}|I^\mu_{\Gamma_0l_0m_0}|^2\, .
\end{equation}

To ensure that our signal was arising only from the HOMO of SO$_2$ and not lower-lying orbitals, we compared the recombination dipole dependence for the HOMO-1 through HOMO-5.  The ionization potentials for sulfur dioxide are 12.35 (HOMO), 12.99 (HOMO-1), 13.22 (HOMO-2), 15.90 (HOMO-3), 16.34 (HOMO-4) and 16.45 (HOMO-5)~\cite{Feng}.  It is thus feasible that we might see some signal from several of the lower-lying orbitals, which are relatively close to the HOMO in energy.  To check this, we plot cross sections for different orbitals and 29.5 eV, the 19th harmonic energy in Fig. S\ref{fig:RecombCrossSec_fig}.  We can see that as we add additional molecular orbitals, the b-axis direction remains dominant at this energy.

\begin{figure}[t]
\centering
\includegraphics[height=100mm]{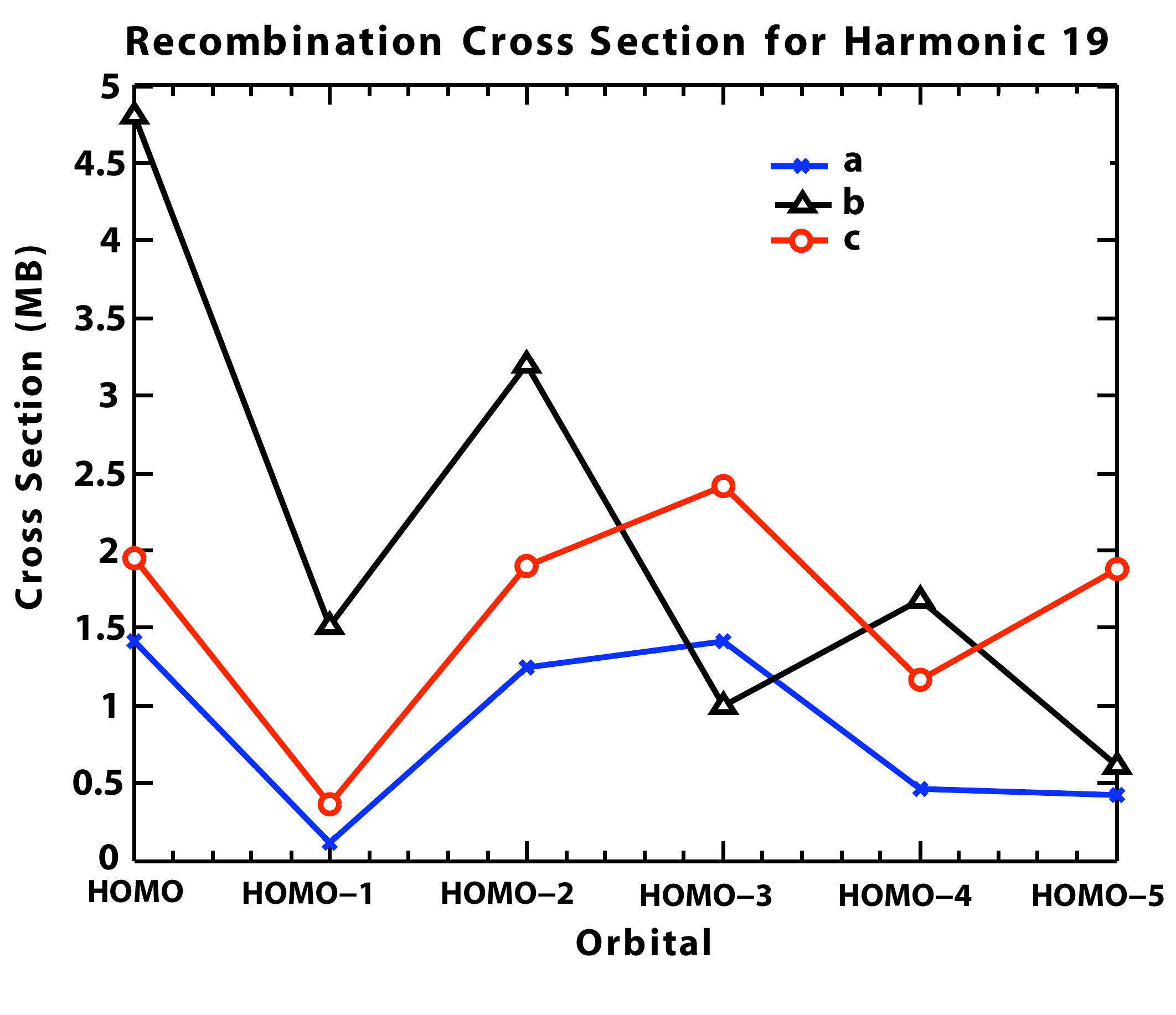}
\caption{Recombination cross section for high harmonic 19 (29.5 eV) at different molecular orbitals.}
\label{fig:RecombCrossSec_fig}
\end{figure}


\end{document}